\documentclass[journal]{IEEEtran}

\usepackage{amsmath}
\usepackage{amssymb}
\usepackage{ams thm}
\usepackage{nccmath}
\usepackage{lipsum}
\usepackage{cite}

\usepackage[pdftex]{graphicx}

\usepackage{graphicx}
\graphicspath{{figs/}}
\DeclareGraphicsExtensions{.pdf,.jpeg,.png,.eps}
\usepackage{eqparbox}
\usepackage{cite}
\usepackage{tabularx}
\usepackage{multirow}
\usepackage[caption=false,font=footnotesize]{subfig}

\usepackage{enumitem}
\usepackage{xcolor}

\usepackage{color,soul}

\newtheorem{postproc}{Algorithm}
\newtheorem{TStype}{Definition}

\begin{document}

\title{Efficient Post-Processors for Improving Error-Correcting Performance of LDPC Codes}

\author{Yaoyu~Tao,~\IEEEmembership{Student~Member,~IEEE,}
	Shuanghong~Sun,~\IEEEmembership{Member,~IEEE,}
	and~Zhengya~Zhang,~\IEEEmembership{Senior~Member,~IEEE}
\thanks{This work was supported in part by NSF CCF-1054270.}
\thanks{Y. Tao, S. Sun, and Z. Zhang are with the Department
of Electrical Engineering and Computer Science, University of Michigan, Ann Arbor, MI, 48109-2122 USA e-mail: (taoyaoyu@umich.edu; shuangsh@umich.edu; zhengya@umich.edu).}
}

\markboth{IEEE Transactions on Circuits and Systems I: Regular Papers}
{Tao \MakeLowercase{\textit{et al.}}: Efficient Post-Processors for Improving Error-Correcting Performance of LDPC Codes}

\maketitle

\begin{abstract}
The error floor phenomenon, associated with iterative decoders, is one of the most significant limitations to the applications of low-density parity-check (LDPC) codes. A variety of techniques from code design to decoder implementation have been proposed to address the error floor problem, among which post-processors have shown to be both effective and implementation-friendly. In this work, we take the inspiration from simulated annealing to generalize the post-processor design using three methods: quenching, extended heating, and focused heating, each of which targets a different error structure. The resulting post-processor is demonstrated to lower the error floors by two orders of magnitude for two structured code examples, a (2209, 1978) array LDPC code, and a (1944, 1620) LDPC code used by the IEEE 802.11n standard. The post-processor can be integrated to a belief-propagation decoder with minimal overhead. The post-processor design is equally applicable to other structured LDPC codes.
\end{abstract}


\IEEEpeerreviewmaketitle

\section{Introduction}

\IEEEPARstart{L}{ow}-density parity-check (LDPC) codes \cite{gallager1962low,mackay1999good} have been widely used in state-of-the-art commercial applications to improve coding gain, measured in the reduction of signal-to-noise ratio (SNR) to meet a required bit error rate (BER) or frame error rate (FER). The coding gain of an LDPC code is captured by a waterfall curve featuring a steep reduction in BER (and FER) with increasing SNR. Popular LDPC codes of block length up to 2 Kb or 4 Kb for wireless \cite{std16e2005,std11n2009,std11ad2012} and wireline applications \cite{std3an2006} have demonstrated excellent waterfall performance down to a BER level of 10\textsuperscript{-7} to 10\textsuperscript{-10}, below which the curve flattens in a phenomenon called error floor \cite{richardson2003error}. The presence of an error floor degrades the achievable BER performance. With future communication and storage systems demanding data rates at multiple Gb/s or higher, error floors will worsen the quality of service. To prevent BER degradation due to error floors, SNR needs to be raised excessively, moving away from the capacity that defines the optimal performance.

Over the last decade, solving the error floor problem has been one research focus in coding theory and decoder design communities. Past experiments have shown that error floors can be caused by practical decoder implementation \cite{zhang2009design}. Improved algorithm implementation and better numerical quantization can suppress these effects \cite{zhang2009design}. However, error floors are fundamentally attributed to noncodeword trapping sets (TS), especially elementary trapping sets (ETS), associated with LDPC codes \cite{mackay2003weaknesses,richardson2003error,landner2005algorithmic}. A TS refers to a set of bits in a codeword, when received incorrectly, cause the belief propagation (BP) decoding algorithm to be trapped in a local minimum \cite{richardson2003error}.

Much work has been done on lowering the error floor by improving code construction using methods such as selective cycle avoidance \cite{tian2004selective}, improved progressive edge growth \cite{xiao2004improved}, code doping \cite{liva2008quasi}, and cyclic lifting \cite{asvadi2011lowering}. These methods are effective, reporting up to 2 orders of magnitudes lower error floor, but they may produce unstructured codes that are not amenable to efficient decoder implementation. The irregular parity check matrices of these techniques complicate the encoder/decoder design, introducing significant overheads in latency, throughput and hardware area. Since theoretical approaches require complete redesign of codes, they are not applicable to the current deployed LDPC systems.

Code concatenation is another approach to lower error floors. With appropriately designed outer codes, such as Bose-Chaudhuri-Hocquenghem (BCH) codes \cite{spourlis2012bch,shieh2015bch} or Reed-Solomon codes \cite{sarika2013rs}, error floors of the concatenated codes can be lowered by up to 2 orders of magnitude. However, the addition of an outer code increases the system complexity, power, cost, and decoding latency.

Alternatively, improvements can be made to decoding using methods such as scaling \cite{cole2006analysis}, averaging \cite{landner2005algorithmic}, and reordering steps \cite{casado2010ldpc} in BP decoding, but the effectiveness of these methods is often limited, usually 1 to 2 orders of magnitude, and some require extra steps that are incompatible with BP decoding, leading to a higher decoder complexity and longer latency. A backtracking approach was proposed in \cite{chen2011hardware} to use a trial and error strategy to flip bits that are likely to be incorrect, and rerun decoding to check if the trial is successful. The approach does not rely on any prior knowledge of trapping sets, but its implementation can be costly in terms of memory and latency. Schedule diversity \cite{lee2014ldpc} was proposed to make multiple decoding attempts using different decoding schedules to reduce the probability of falling into a trapping set. The approach is another form of trial and error, and can be costly in latency.

In theory, a more effective approach is to add a post-processing step if a decoding error is detected, and the post-processing is done in a targeted manner without having to rely on trial and error. An example of post-processing is the bi-mode syndrome-erasure decoding algorithm \cite{han2009low,zhang2008lowering}. One drawback of the post-processing approach is that it is usually limited to specific codes and it is not known whether it is generally applicable. A redecoding approach based on attenuating a predetermined set of bits \cite{lee2018effective} was proposed for quasi-cyclic (QC) LDPC codes. The approach involves an offline search and it may only be applied to QC-LDPC codes.

\begin{table*}
	\centering
	\caption{Comparison of Techniques for Lowering LDPC Error Floors}
	\renewcommand{\arraystretch}{1.2}
	\begin{tabular}{|c|c|c|c|c|c|c|c|c|}
		\hline
		\multirow{2}{*}{Techniques} & Cycle avoidance \cite{tian2004selective}, PEG\cite{xiao2004improved}, & LDPC+ & LDPC+ & Scaled  & Reordered & Bi-mode & Postprocessing\\
		& cyclic lift \cite{asvadi2011lowering}, code doping \cite{liva2008quasi} & BCH \cite{spourlis2012bch} & RS \cite{sarika2013rs} & BP \cite{cole2006analysis} & BP \cite{casado2010ldpc} & \cite{han2009low} & (this work)\\ \hline\hline
		\multirow{2}{*}{LDPC code} & \multirow{2}{*}{irregular} & (504,252) & (1944,972) & (2640,1320) & (2016,1512) & (1984,1240) & (2209,1978)\\
		 &  & regular & regular & regular & regular & regular & regular \\ \hline
		\multirow{2}{*}{Outer code} & \multirow{2}{*}{---} & (756,696,6) & (155,75) & \multirow{2}{*}{---} & \multirow{2}{*}{---} & \multirow{2}{*}{---} & \multirow{2}{*}{---} \\
		 &  & BCH code & RS code &  &  &  & \\ \hline
		Error floor & \multirow{2}{*}{$\sim$10$\times$} & \multirow{2}{*}{$\sim$100$\times$} & \multirow{2}{*}{---} & \multirow{2}{*}{10$\sim$100$\times$} & \multirow{2}{*}{$\sim$10$\times$} & \multirow{2}{*}{100$\sim$1000$\times$} & \multirow{2}{*}{100$\sim$1000$\times$}\\
		lower by &  &  &  &  &  &  & \\ \hline
		\multirow{2}{*}{Latency} & high (due to & high (due to & high (due to & \multirow{2}{*}{negligible} & \multirow{2}{*}{moderate} & \multirow{2}{*}{moderate} & \multirow{2}{*}{negligible}\\
		& code irregularity)  & outer code) & outer code) &  &  &  & \\ \hline
		\multirow{2}{*}{Code rate loss} & \multirow{2}{*}{no loss} & \multirow{2}{*}{0.75$\to$0.69} & \multirow{2}{*}{0.62$\to$0.30} & \multirow{2}{*}{no loss} & \multirow{2}{*}{no loss} & \multirow{2}{*}{negligible} & \multirow{2}{*}{no loss}\\
		&  &  &  &  &  &  & \\ \hline
		\multirow{2}{*}{Hardware cost} & high (due to & high (extra & high (extra & \multirow{2}{*}{negligible} & \multirow{2}{*}{negligible} & \multirow{2}{*}{low} & negligible\\
		& code irregularity) & BCH block) & RS block) &  &  &  & (Table~\ref{tbl:hardware})\\  \hline
		\multirow{2}{*}{Throughput loss} & high (due to & \multirow{2}{*}{negligible} & \multirow{2}{*}{negligible} & \multirow{2}{*}{negligible} & \multirow{2}{*}{negligible} & \multirow{2}{*}{negligible} & \multirow{2}{*}{negligible} \\
		& code irregularity) &  &  &  &  &  & \\ \hline
		Back compatibility & no & no & no & yes & yes & yes & yes\\ \hline
	\end{tabular}
	\label{tbl:errorfloorcmp}
\end{table*}

In this work, we extend the post-processing method that was first presented in \cite{zhang2008lowering} using ideas from simulated annealing (SA). The SA algorithm combines random walk (or heating in annealing terminology) and gradient descent (or cooling) to escape local minima \cite{kirkpatrick1983optimization,vcerny1985thermodynamical,reeves1993modern}. In post-processing, we use message reweighting or soft bit flipping to perturb, or heat up, local minima, and use BP to cool down for convergence towards a codeword. Compared to well-known approaches above, the cost of implementing post-processing is low: no code change is needed, and the post-processing is entirely based on BP. As post-processing is conditionally invoked, i.e., when a decoder fails to converge at a low BER, the impact on decoding throughput and power is negligible. 

A preliminary form of post-processing was first shown in \cite{zhang2008lowering}. We call this form of post-processing ``quenching'', referring to one iteration of heating followed by immediate cooling. Quenching was demonstrated to lower the error floor of a (2048, 1723) RS-LDPC code from a BER level of 10\textsuperscript{-10} to 10\textsuperscript{-14} by removing the vast majority of the errors until only the minimum distance errors remain \cite{zhang2010efficient}, but the quenching method is not as effective in other structured LDPC codes.

In this paper, we present two new methods inspired by SA: extended heating and focused heating, aiming at eliminating the vast majority of ETS errors of various structures. We use a rate-0.89 (2209, 1978) array LDPC code \cite{fan2000array} that is known for its collection of ETS errors \cite{zhang2009design} to derive these methods. Finally, we combine extended heating and focused heating into a generalized method that is applicable to LDPC codes with unknown ETS structures. We use a rate-0.83 (1944, 1620) LDPC code for the IEEE 802.11n standard \cite{std11n2009} to test the effectiveness of the generalized method. Experimental results show that post-processing is one of the most practical and efficient solutions in designing low-error-floor LDPC decoders. Table~\ref{tbl:errorfloorcmp} summarizes the qualitative features of the proposed approach compared to prior techniques.

\section{Background}

An LDPC code is defined by a sparse $m \times n$ parity-check matrix $H$ where $n$ represents the number of bits in the code block and $m$ represents the number of parity checks \cite{gallager1962low,mackay1999good}. The $H$ matrix of an LDPC code can be illustrated graphically using a bipartite graph, where each bit is represented by a variable node (VN) and each parity check is represented by a check node (CN). VN is also called bit, and CN is also called check or parity check. An edge exists between VN $i$ and CN $j$ if and only if $H(j,i) = 1$.

An LDPC code is decoded using the BP algorithm that operates on factor graphs \cite{mackay1999good}. Soft messages representing reliabilities are exchanged between VNs and CNs to compute the likelihood of whether a bit is 1 or 0. The BP algorithm has two popular implementations, the sum-product algorithm and the min-sum algorithm \cite{hagenauer1996iterative,fossorier1999reduced}. The min-sum algorithm is simpler to implement and provides excellent decoding performance with suitable corrections \cite{chen2005reduced}. It has been widely used in hardware decoders. In this work, we will base our discussions on the min-sum algorithm due to its practical relevance.

\subsection{Min-Sum Decoding}

Assume a binary phase-shift keying (BPSK) modulation and an additive white Gaussian noise (AWGN) channel. The binary values 0 and 1 are mapped to 1 and -1, respectively. The min-sum decoding can be explained using the factor graph. In the first step of decoding, each VN $x_i$ is initialized with the prior log-likelihood ratio (LLR) defined in \eqref{eqn_prior} based on the channel output $y_i$:

\begin{equation}
\label{eqn_prior}
L^{pr}(x_i) = \log{\frac{\Pr{(x_i = 0 \mid y_i)}}{\Pr{(x_i = 1 \mid y_i)}}} = \frac{2}{\sigma^2} y_i
\end{equation}

\noindent where $\sigma^2$ represents the channel noise variance.

After initialization, VNs send the prior LLRs to the CNs along the edges defined by the factor graph. The LLRs are recomputed based on parity checks, as in equation \eqref{eqn_c2v}, and returned to the VNs. Each VN then updates its decision based on the posterior LLR that is computed as the sum of the prior LLR from the channel and the LLRs received from the CNs, as in equation \eqref{eqn_post}. One round of message exchange between VNs and CNs completes one iteration of decoding. To start the next iteration, each VN computes the marginalized LLRs, as in equation \eqref{eqn_v2c}, and passes them to the CN.

\begin{equation}
\label{eqn_c2v}
L(r_{ij}) = \min_{i' \in Row[j] \setminus i} \left| L(q_{i'j}) \right| \prod_{i' \in Row[j] \setminus i} \operatorname{sgn} \left( {L(q_{i'j})} \right)
\end{equation}

\begin{equation}
\label{eqn_post}
L^{ps}(x_i) = \sum_{j' \in Col[i]} {L(r_{ij'})} + L^{pr}(x_i)
\end{equation}

\begin{equation}
\label{eqn_v2c}
L(q_{ij}) = L^{ps}(x_i) - L(r_{ij})
\end{equation}

The LLRs passed between VNs and CNs are known as the variable-to-check message (VC message, $L(q_{ij})$) and check-to-variable message (CV message, $L(r_{ij})$), where $i$ is the VN index and $j$ is the CN index. In representing the connectivity of the factor graph, $Col[i]$ refers to the set of all the CNs connected to the $i$th VN and $Row[j]$ refers to the set of all the VNs connected to the $j$th CN.

The magnitude of $L(r_{ij})$ computed using \eqref{eqn_c2v} is overestimated and correction terms are introduced to reduce the approximation error. The correction is in the form of either an offset or a normalization factor \cite{chen2005reduced}.

A hard decision is made in each iteration based on the posterior LLR, as in \eqref{eqn_hard}. The iterative decoding is allowed to run until the hard decisions satisfy all the parity checks or when an upper limit on the iteration number is reached.

\begin{equation}
\label{eqn_hard}
\hat{x_i} = \begin{cases}
0& \text{if $L^{ps}(x_i) \geq 0$}\\
1& \text{if $L^{ps}(x_i) < 0$}
\end{cases}
\end{equation}

In a practical decoder implementation, the VC messages and CV messages are quantized to fixed point. We use the notation $Qp.q$ to indicate a two's-complement fixed-point quantization with $p$ bits for integer and $q$ bits for fraction.

\subsection{Error Floor and Trapping Set}

It is known that TS is the fundamental cause of error floor in BP decoding of LDPC codes \cite{richardson2003error}. We repeat the definition of TS \cite{landner2005algorithmic} and a special type of TS called elementary TS, or ETS, that is the most dominant in error floors.

\begin{TStype}
\label{def_TS}
Trapping set (TS) and elementary trapping set (ETS)

An $(a,b)$ TS is a configuration of $a$ number of VNs, for which the induced subgraph in $G$ contains $b>0$ odd-degree CNs with respect to the TS. An $(a,b)$ ETS is a TS for which all CNs in the induced subgraph have either degree 1 or 2 with respect to the TS, and there are exactly $b$ CNs of degree 1 with respect to the TS. The CNs of degree 1 are called degree-1 CNs, and the CNs of degree 2 are called degree-2 CNs.

\end{TStype}

\begin{figure}
\centering
\includegraphics[width=\linewidth]{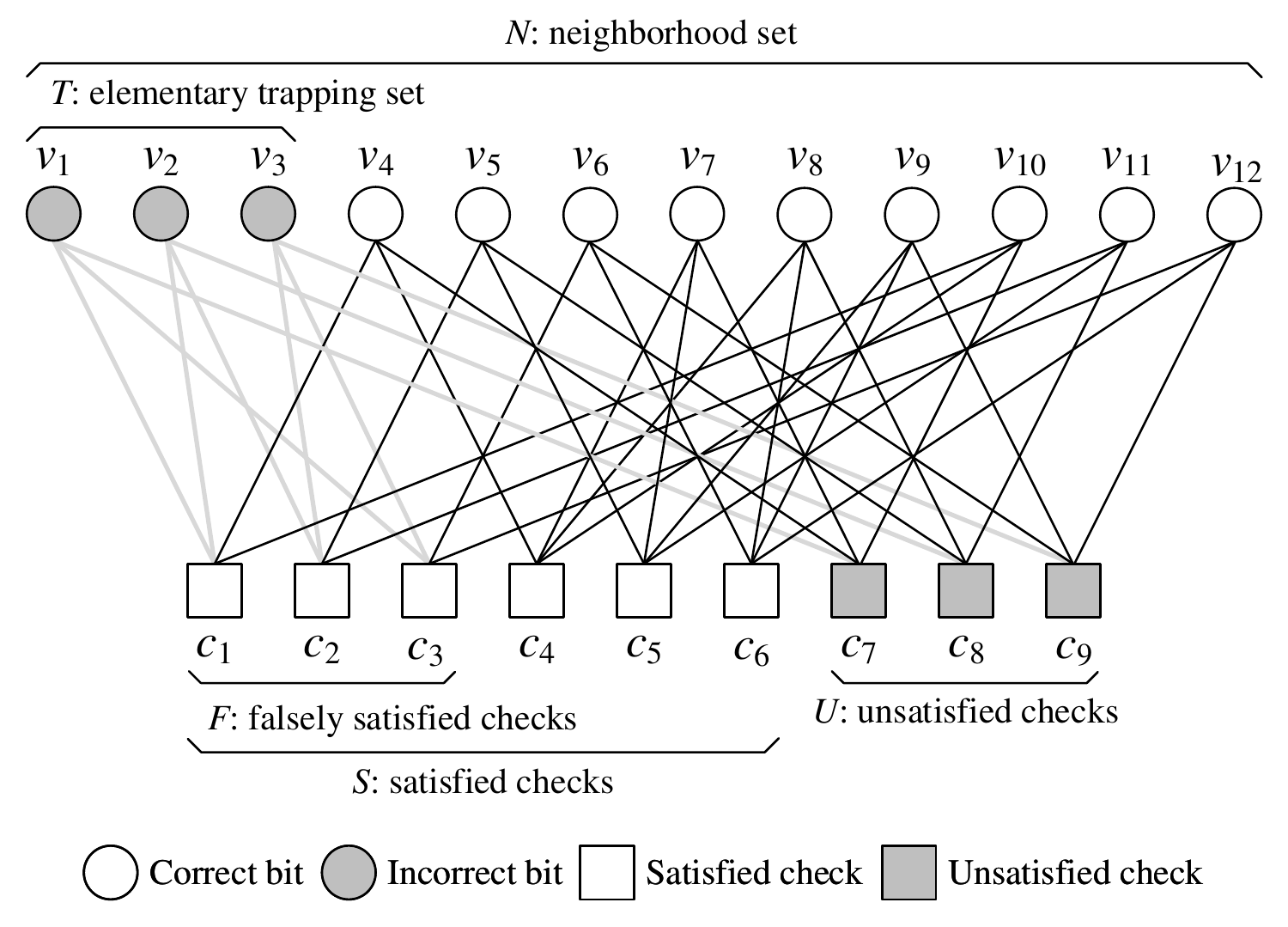}
\caption{Illustration of a (3,3) ETS.}
\label{absorbing_set}
\end{figure}

We will focus the following discussions on ETS as it is the most common type of TS and the most damaging in causing error floors. A factor graph of a small LDPC code is shown in \figurename~\ref{absorbing_set}, which contains a (3,3) ETS $\mathcal{T}$. Each VN in $\mathcal{T}$ is connected to 1 degree-1 CN and 2 degree-2 CNs.

To see how an ETS can cause a decoding error, we use an example of transmitting an all-zero vector of length 12, which is a codeword for the code defined in \figurename~\ref{absorbing_set}. Suppose the received word contains errors in the first three bits. That is, the VNs in $\mathcal{T}$ are initialized to 1 and the remaining VNs are initialized to 0. The received word does not constitute a valid codeword, as the degree-1 CNs labeled $\mathcal{U}$ are not satisfied. Note that among the satisfied CNs labeled $\mathcal{S}$, the degree-2 CNs labeled $\mathcal{F}$ are falsely satisfied, i.e., the ones that are connected to an even number of bits in $\mathcal{T}$. In BP decoding, the CV messages from the degree-1 CNs in $\mathcal{U}$ will attempt to correct the wrong bits, but the CV messages from the degree-2 CNs in $\mathcal{F}$ will reinforce the wrong bits. If there is stronger reinforcement than correction of the wrong bits, the decoder is trapped in the non-codeword ETS.

Many LDPC codes contain ETS of lower weight than the minimum distance of the code. As a result, the decoders can be more easily trapped in an ETS at a moderate to high SNR level than converging to a minimum-distance codeword. The presence of ETS results in error floors. Reducing the likelihood of trapping in the local minimum due to ETS is the key to lowering the error floors of LDPC codes.

\section{Simulated Annealing and Post-processing in BP Decoding}

The local minimum problem has been studied extensively in the field of optimization. Notably, the SA algorithm combines gradient descent and random walk to escape local minima \cite{kirkpatrick1983optimization,vcerny1985thermodynamical,reeves1993modern}. Annealing is a process in metallurgy, where metal is heated to a high temperature and then undergoes controlled cooling to form a low-energy crystalline structure. If metal contains no defects, its energy is at the minimum; otherwise, it will be at a higher energy level. An analogy can be made for decoding: the highest energy occurs in the beginning of decoding when most errors or defects are present. As decoding proceeds, errors are corrected and the energy goes down, just as the cooling process in annealing that removes defects. When the decoding converges to a correct codeword, the energy goes down to the minimum, like metal reaching its defect-free, lowest-energy crystalline state.

The decoder can be trapped in an ETS. The weight of the ETS that induce error floors is often lower than the minimum distance. If an ETS is within the minimum distance away from the correct codeword, a local search algorithm can be applied. SA is such an algorithm that targets local minimum problems.

\subsection{Neighborhood Identification for Trapping Sets}

SA uses heating to perturb the local minimum, making it unstable before breaking away from it. The most efficient way is to heat only the defective points in order to keep the amount of perturbation low and reduce the risk of moving much further away from the closest global minimum. Similarly in LDPC decoding, heating needs to be directed to the error bits in an ETS. The ETS is not known, but the degree-1 CNs are known because they are not satisfied. We can trace the neighboring VNs of the degree-1 CNs, called the neighborhood set $\mathcal{N}$, as labeled in \figurename~\ref{absorbing_set}.

The neighborhood set contains one or more VNs in the ETS, and also VNs outside of the ETS. There is no choice but to apply heating to the entire neighborhood set. As a result, heating will perturb not only the error bits but also the correct bits. In practice, the neighborhood set can be as large as tens or a few hundred bits, therefore heating needs to be carefully adjusted to be effective to resolve the local minimum, but not too much to be pushed to a different codeword.

\subsection{Heating}

Heating is used to perturb the local minimum. In BP decoding, perturbation can be done by reweighting the VC and CV messages \cite{zhang2008lowering} or soft bit flipping. In \figurename~\ref{absorbing_set}, the bits in the ETS $\mathcal{T}$ are incorrect. Each VN in $\mathcal{T}$ receives CV messages from 2 degree-2 CNs to reinforce the error and a CV message from 1 degree-1 CN that attempts to correct the error. To perturb this local minimum and possibly escape the local minimum, the CV messages from the satisfied CNs (including degree-2 CNs) are weakened, and the CV message from the degree-1 CN is strengthened. This procedure is called message reweighting. As the magnitude of the messages are changed, noise is injected to the system to achieve a perturbation effect.

Message reweighting applied to the VNs in an ETS helps correct errors, but message reweighting applied to the VNs outside the ETS can possibly introduce more errors. For example, in \figurename~\ref{absorbing_set}, $v_4 \notin \mathcal{T}$, and $v_4$ is connected to $c_1$ and $c_4$ that are both satisfied and $c_7$ that is degree-1 and unsatisfied. By the reweighting procedure outlined above, the CV messages from $c_1$ and $c_4$ to $v_4$ are weakened, and the CV message from $c_7$ to $v_4$ is strengthened, which is likely to cause $v_4$ to flip to the incorrect value. Therefore, heating needs to be carefully adjusted to avoid perturbing too many correct bits and eventually converge to an undesired global minimum.

\subsection{Post-Processing Procedure}

BP decoding with post-processing follows a two-phase procedure. In the first phase, conventional BP decoding is performed. If BP decoding fails to converge after a set number of iterations, denoted as $M$, at a moderate to high SNR, the decoding is most likely trapped in a local minimum and it enters the second phase.

In the second phase, post-processing is invoked. Neighborhood set needs to be properly identified for effective heating. The identification can be conveniently done in VN by inspecting the sign of incoming CV messages: if the sign indicates that the parity check is unsatisfied, the VN tags the bit as part of the neighborhood set $\mathcal{N}$. Heating is performed by reweighting the reliability of CV messages, i.e., increasing the reliability of CV messages from the unsatisfied checks to $\mathcal{N}$, or decreasing the reliability of CV messages from the satisfied checks to $\mathcal{N}$, or both. Equation {\eqref{heating_eq}} describes a way to implement message reweighting that decreases the reliability of the CV messages from the satisfied CNs $\mathcal{S}$ to the VNs in the neighborhood set $\mathcal{N}$ to a low value $A_0$. The value of $A_0$ determines the amount of heating, or perturbation injected to the local minimum. Heating can also be done using soft bit flipping to be described in Section~\ref{sec:type3}.

\begin{equation}
\begin{gathered}
\begin{split}
\label{heating_eq}
L(r_{ij}) = \prod_{i' \in Row[j] \setminus i} \operatorname{sgn} \left( L(q_{i'j}) \right) \cdot \\
\begin{cases}
A_0 & \text{if $v_i\in\mathcal{N}$, $c_j\in \mathcal{S}$}\\
\;\;\min_{i' \in Row[j] \setminus i} \left| L(q_{i'j}) \right| & \text{otherwise}.
\end{cases}
\end{split}
\end{gathered}
\end{equation}

After $P$ iterations of heating, $N$ iterations of BP decoding is applied to cool down. The post-processing procedure is summarized in Algorithm~\ref{alg_pp}.

\begin{postproc}
\label{alg_pp}
Post-Processing Procedure
	\begin{enumerate}
		\item BP decoding: run for $M$ iterations. If there are unsatisfied CNs, continue post-processing.
		\item Post-processing:
		\begin{enumerate}
			\item Heating: run $P$ iterations of reweighted message passing.
			\item Cooling: run $N$ iterations of BP decoding.
		\end{enumerate}
	\end{enumerate}
\end{postproc}

In Algorithm~{\ref{alg_pp}}, $M$ is set to ensure that the decoder has been trapped in an ETS, and $N$ is set to ensure that the decoder has enough time to cool down to the global minimum after heating. In this paper, we set $M = N = 20$.

\subsection{Implementing Post-Processing in Hardware}
\label{sec:hardware}

The primary design goal of post-processing is to lower the error floor with minimal cost of area, power, latency and throughput. An ideal post-processor works likes a ``plug-in'' feature that can be easily integrated to any standard LDPC decoder.

In a standard min-sum LDPC decoder, a CN is implemented as a comparison tree to find the first and the second minimum. A CN often contains little memory and does not retain states. On the other hand, a VN keeps state and stores prior and posterior information. If post-processing is implemented by reweighting CV messages, a CN needs to be augmented to keep track of all the VNs in the neighborhood set $\mathcal{N}$, which could be costly. Therefore, instead of reweighting CV messages, we devise an alternative by reweighting VC messages. In this alternative approach, a VN is augmented by 1 bit to track whether it belongs to the neighborhood set $\mathcal{N}$. Because the magnitude of VC messages tends to saturate to the maximum value allowed by quantization in a few iterations, the reweighting is implemented in VN by decreasing the magnitude of the VC message from a VN in the neighborhood set to a satisfied CN to a low value $A_0$. The reweighted (magnitude-reduced) VC message propagates to the satisfied CN, and through the CN's minimum operation becomes reweighted CV message. Equation {\eqref{heating_eq_vc}} describes post-processing by reweighting VC messages.

\begin{equation}
\begin{gathered}
\begin{split}
\label{heating_eq_vc}
L(q_{ij}) = 
\begin{cases}
A_0\cdot{sgn(L^{ps}(x_i) - L(r_{ij}))} & \text{if $v_i\in\mathcal{N}$, $c_j\in \mathcal{S}$}\\
L^{ps}(x_i) - L(r_{ij}) & \text{otherwise}.
\end{cases}
\end{split}
\end{gathered}
\end{equation}

The choice of $A_0$ depends on the quantization. Assume VC messages are quantized to $Qp.q$, the possible $A_0$ values are $\{0, 2^{-q}, 2^{-q+1}, ..., 2^{p-1}-2^{-q}\}$. The lower the $A_0$, the more noise is injected to the local minimum. As a result, lower $A_0$ is more effective in resolving an ETS error, but also highly likely to cause more perturbation to the bits outside the ETS, which may push the decoder to an undesired global minimum. Detailed message reweighting strategies are dependent on the structures of the ETS, which will be elaborated in Section IV.

Post-processing does not require changing the code structure or decoder architecture. Muxes and label bit registers are added for VC message reweighting and neighborhood identification, respectively. A controller monitors the decoding and enables post-processing upon detecting failed CNs after $M$ iterations; therefore post-processing is activated at a rate of approximately the decoding FER and has a negligible impact on the decoding throughput and the average latency.

\begin{figure}
\centering
\includegraphics[width=\linewidth]{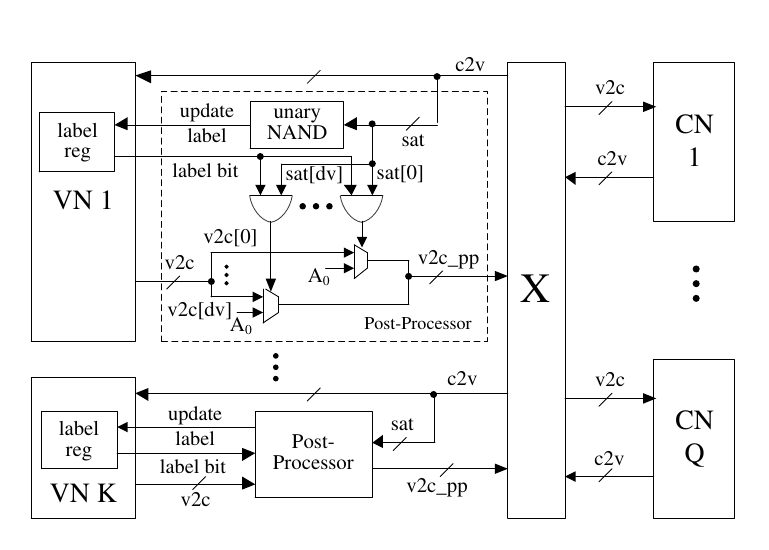}
\caption{Post-processing added to a fully-parallel decoder.}
\label{pp_fully_parallel}
\end{figure}

We demonstrate post-processing implementation based on two commonly used LDPC decoder architectures, the fully-parallel architecture \cite{fully_par_ldpc} and the row-parallel architecture \cite{blanksby_ldpc}. A fully-parallel architecture is efficient for short code length and it yields the highest throughput. In a fully-parallel decoder, all CNs, VNs and their interconnections are instantiated in hardware exactly as those in the code's factor graph. Assume a decoder contains $Q$ CNs and $K$ VNs, and the VN degree is $d_v$. At each VN, a label register is added to indicate whether the VN belongs to the neighborhood set, and a post-processor is added to perform neighborhood labeling and VC message reweighting, as shown in {\figurename~\ref{pp_fully_parallel}}. The post-processor takes the signs of $d_v$ CV messages (without marginalization) as inputs $sat$ to identify whether CNs are satisfied. If post-processing is enabled and at least one incoming CV message indicates that the CN is unsatisfied, the post-processor turns the VN's neighborhood label on with a unary NAND gate. In performing post-processing, the VN's neighborhood label is AND'ed with the $sat$ of each CV message to determine whether reweighting is enabled. If reweighting is enabled, a MUX is used to select the reduced magnitude of $A_0$ for the outgoing VC message $v2c\_pp$.

\begin{figure}
\centering
\includegraphics[width=\linewidth]{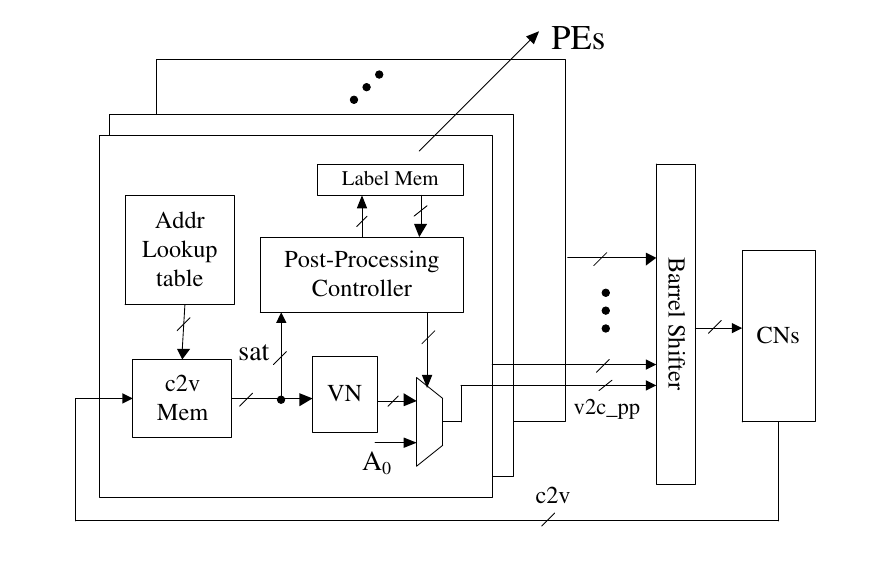}
\caption{Post-processing added to a row-parallel decoder.}
\label{pp_row_parallel}
\end{figure}

A row-parallel decoder architecture is the most popular architecture for moderate to long QC LDPC codes, including many that have been used in standards. A row-parallel architecture often employs layered BP decoding. Each iteration is divided into multiple layers of processing. An example decoder architecture is shown in \figurename~\ref{pp_row_parallel}. Each layer processing is done by multiple processing elements (PEs), each consisting of a physical VN and memory. The read/write addresses are stored in lookup tables. In the row-parallel architecture, a physical VN is time-multiplexed and it assumes the roles of multiple logical VNs (VNs in the factor graph), one in each layer. A label memory is added to store the neighborhood labels of the logical VNs. A post-processing controller is added to the PE to perform the same labeling and reweighting functions as what the post-processor does in the fully-parallel architecture. The only difference is that the post-processing controller performs the labeling and reweighting serially as the CV messages are received one at a time.

Table~\ref{tbl:pp_impl} shows the overhead when post-processing is added to a fully-parallel decoder and a row-parallel decoder for the IEEE 802.11n (648,540) LDPC code and the IEEE 802.11n (1944,1620) LDPC code, respectively. The percentage in the brackets indicate the device utilization. The 100 MHz clock frequency can be kept even after post-processing is added, so the average throughput and latency can be kept constant. Implementing post-processing on the row-parallel decoder uses 4.5\% and 8.3\% more slice registers and slice LUTs, respectively, compared to the baseline. The cost is even lower when post-processing is added to the fully-parellel decoder.

\begin{table}
	\centering
	\caption{Evaluation of Post-Processing Implementations of Fully-Parallel and Row-Parallel Decoders (based on Xilinx Virtex-5 XC5VLX155T FPGA)}
	\renewcommand{\arraystretch}{1.2}
	\begin{tabular}{|c|c|c|c|c|c|c|}
		\hline
		\multirow{2}{*}{Design} & \multicolumn{2}{c|}{Fully-par. (648,540) dec.} & \multicolumn{2}{c|}{Row-par. (1944,1620) dec.} \\
		\cline{2-5}
		& Baseline & Post-proc added & Baseline & Post-proc added \\ \hline\hline
		Slice & 13,724 & 13,901 & 4,432 & 4,633 \\ 
		registers & (14.24\%) & (14.43\%) &  (4.60\%) & (4.81\%) \\\hline
		Slice & 39,007 & 40,822 & 10,066 & 10,901\\ 
		LUTs & (40.30\%) & (42.17\%) & (10.4\%) & (11.2\%)\\\hline
		Occupied & 10,852 & 11,208 & 4,782 & 4,844\\ 
		slices & (44.70\%) & (46.17\%) & (19.7\%) & (19.9\%)\\\hline
		\multirow{2}{*}{BRAMs}  & 64 & 64 & 35 & 35\\
		& (29.9\%) & (29.9\%) & (16.4\%) & (16.4\%)\\\hline
	\end{tabular}
	\label{tbl:pp_impl}
\end{table}

\section{Error Structure and Post-Processing Methods}

Studying the error floor phenomenon requires fast simulations. FPGA accelerated emulations are particularly useful because software-based simulations often take weeks or months to reach low BER levels. In previous work \cite{li2012reconfigurable}, a library and script based approach was developed to automate the FPGA emulations for LDPC decoders. In this work, we used it to collect errors in the error floor region.

After collecting enough errors in the error floor region, we analyze the ETS structures associated with these errors. The ETS structures are dependent on the code structure. The post-processing method is formulated to be the most effective towards the structures.

\subsection{Type I ETS and Quenching}

The (2048,1723) RS-LDPC code \cite{djurdjevic2003class} for the IEEE 802.3an standard \cite{std3an2006} is a well-studied code for error floor investigation  \cite{zhang2009design}. The $H$ matrix of this regular code has a column degree of 6, a row degree of 32, and 64$\times$64 permutation matrices as component submatrices \cite{djurdjevic2003class}. The code has a girth of at least 6. The code has an error floor below 10\textsuperscript{-10}. It has been shown that the error floor is dominated by (8,8) ETS errors \cite{zhang2009design}.

\begin{figure}
\centering
\includegraphics[width=.75\linewidth]{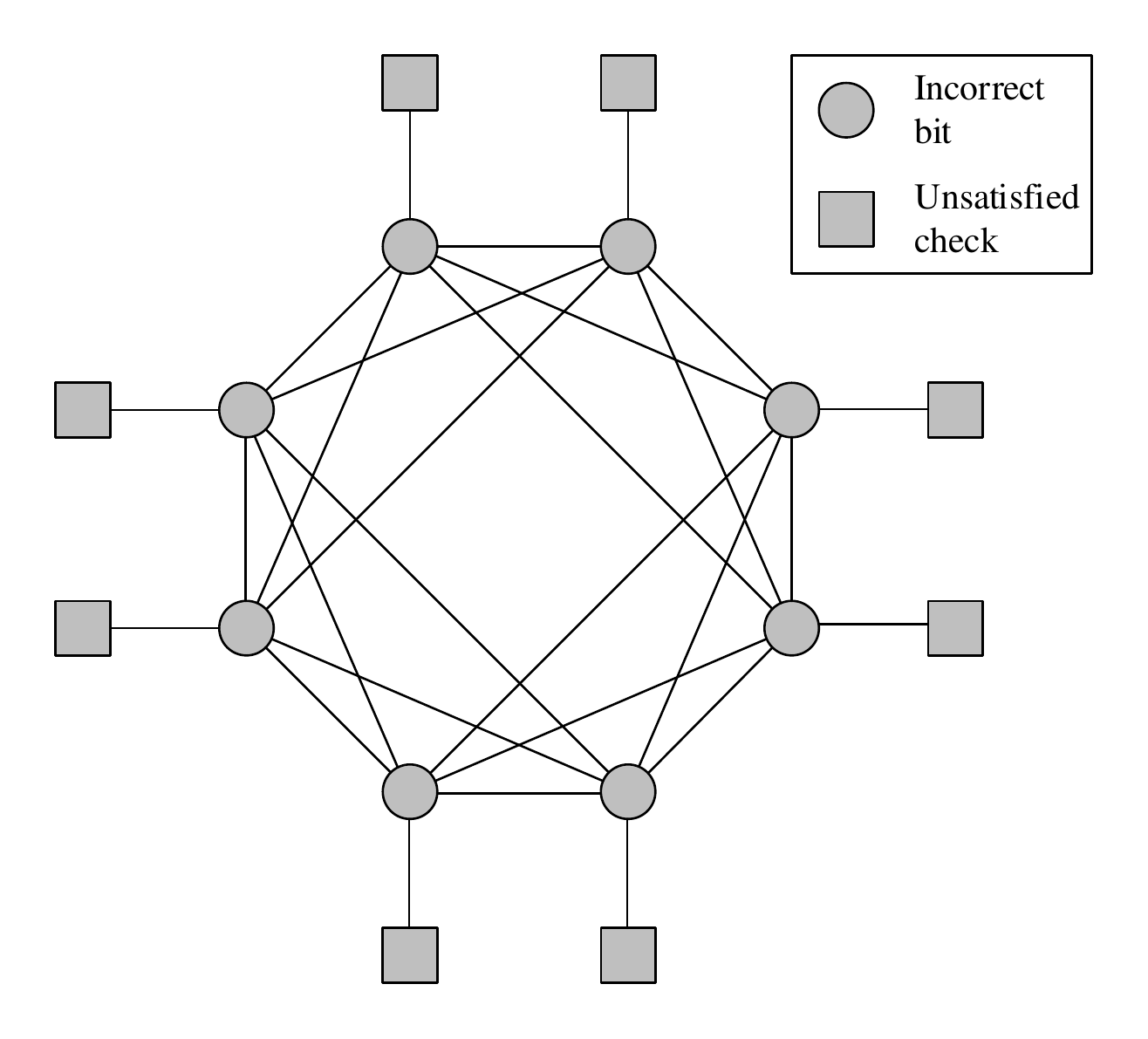}
\caption{An (8,8) ETS of a (2048,1732) RS-LDPC code.}
\label{rs_88}
\end{figure}

The (8,8) ETS is illustrated in \figurename~\ref{rs_88} using a simplified representation that only includes VNs in the ETS and degree-1 CNs. The (8,8) ETS consists of 8 VNs, each of which is connected to one degree-1 CN. The degree-2 CNs are shown implicitly in \figurename~\ref{rs_88} as lines connecting pairs of VNs in the ETS. The illustration makes it clear if the bits in the ETS are initialized with incorrect binary values, these VNs will reinforce each other through the degree-2 CNs. As each VN in the ETS neighbors 5 degree-2 CNs and only 1 degree-1 CN, a BP decoder can be easily trapped in this local minimum. The (8,8) ETS is an example of a Type I ETS. \textbf{A Type I ETS is one in which each VN is connected to exactly 1 degree-1 CN.}

To resolve a Type I ETS error, Algorithm~\ref{alg_pp} can be used with $P=1$, i.e., only one iteration of heating followed by immediate cooling, as proposed by \cite{zhang2008lowering}. This post-processing method is named ``quenching''. Quenching is effective towards Type I ETS errors, since the neighborhood set traced from the unsatisfied degree-1 CNs contains the entire ETS. One iteration of heating reaches all VNs in the ETS, and cooling can be applied immediately after to help convergence.

Using the (8,8) ETS illustrated in \figurename~\ref{rs_88} as an example, after the heating step, each VN in the ETS receives 5 weakened CV messages from degree-2 CNs and 1 CV message from a degree-1 CN. The lower the reweighted value $A_0$ is, the more likely the CV message from the degree-1 CN can overcome the sum of 5 weakened CV messages from the degree-2 CNs.

\begin{figure}
\centering
\includegraphics[width=.75\linewidth]{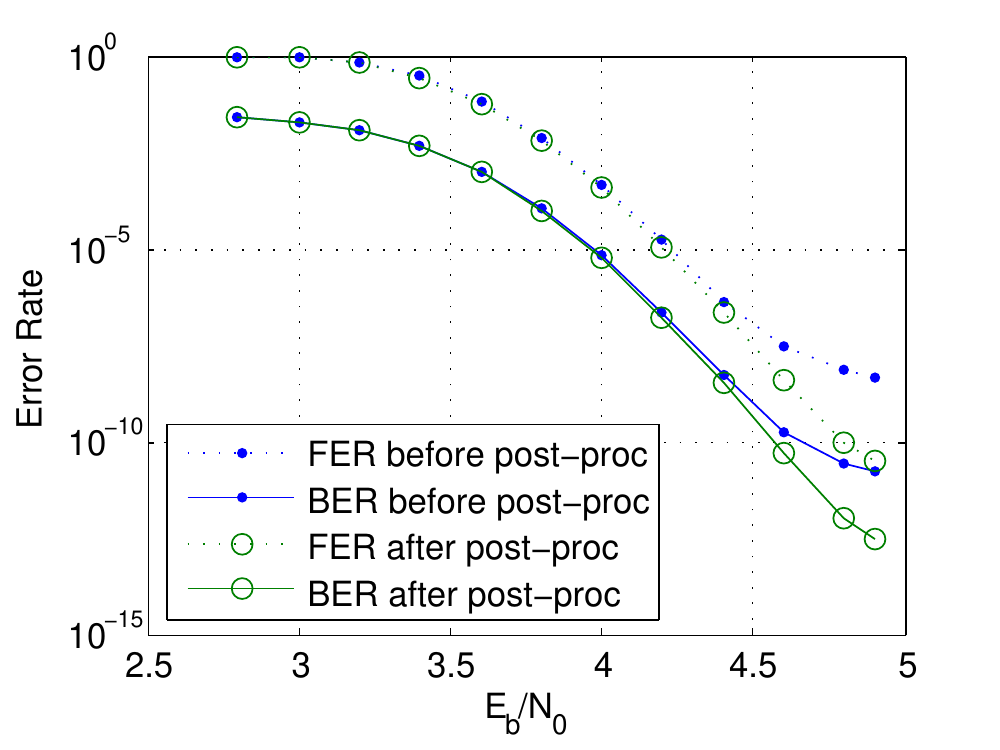}
\caption{Error rate of the (2048, 1723) RS-LDPC code before and after post-processing using quenching \cite{zhang2008lowering}.}
\label{rs_ber}
\end{figure}

Previous work showed that over 97\% of the ETS errors in the error floor region of the (2048, 1723) RS-LDPC code are corrected using quenching with proper choice of $A_0$, resulting in nearly two orders of magnitude lower error floor as shown in \figurename~\ref{rs_ber} \cite{zhang2008lowering}. $A_0$ = 1 is used in this experiment.

\subsection{Type II ETS and Extended Heating}

To extend from previously proposed quenching post-processing \cite{zhang2008lowering}, we choose a (5,47)-regular rate-0.89 (2209, 1978) array LDPC codes \cite{fan2000array} for investigation of other types of ETS structures. The $H$ matrix of this code can be partitioned into 5 row groups and 47 columns groups of 47$\times$47 permutation matrices. 

We collected 274 errors through FPGA emulation of a $Q4.0$ (2209, 1978) array LDPC decoder in the error floor region ($E_b/N_0$ = 5.6 dB, 5.8 dB, and 6.0 dB). 243 out of the total 274 errors are ETS errors, among which there is only 1 Type I ETS error. We then applied quenching to post-process these errors and the results are summarized in Table~\ref{tbl:quenchingarray}. A resolving rate of only 73\% indicates that quenching alone is not sufficient to lower the error floor of array code.

\begin{table}
	\centering
	\caption{Error Profile of the (2209, 1978) Array LDPC Code and Effectiveness of Quenching}
	\renewcommand{\arraystretch}{1.2}
	\begin{tabular}{|c|c|c|}
		\hline
		\multirow{2}{*}{ETS} & Error & Resolved by\\
		& count & quenching \\ \hline
		(6,8)  & 6 & 3 (50\%)\\ \hline
		(7,9)  & 5 & 2 (40\%)\\ \hline
		(8,6)  & 124 & 105 (85\%)\\ \hline
		(8,8)  & 20  & 13 (65\%)\\ \hline
		(9,5)  &  37 & 33 (89\%)\\ \hline
		(10,4) &  12 &  8 (67\%)\\ \hline
		(10,6) &   9 &  5 (56\%)\\ \hline	
		(10,8) &   7 &  7 (100\%)\\ \hline
		other ETS  & 23 &  12 (52\%)\\ \hline\hline
		non-ETS &  31 & 11 (58\%)\\ \hline\hline
		Total  & 274 & 199 (73\%)\\ \hline
	\end{tabular}
	\label{tbl:quenchingarray}
\end{table}

\begin{figure}
\centering
\includegraphics[width=.75\linewidth]{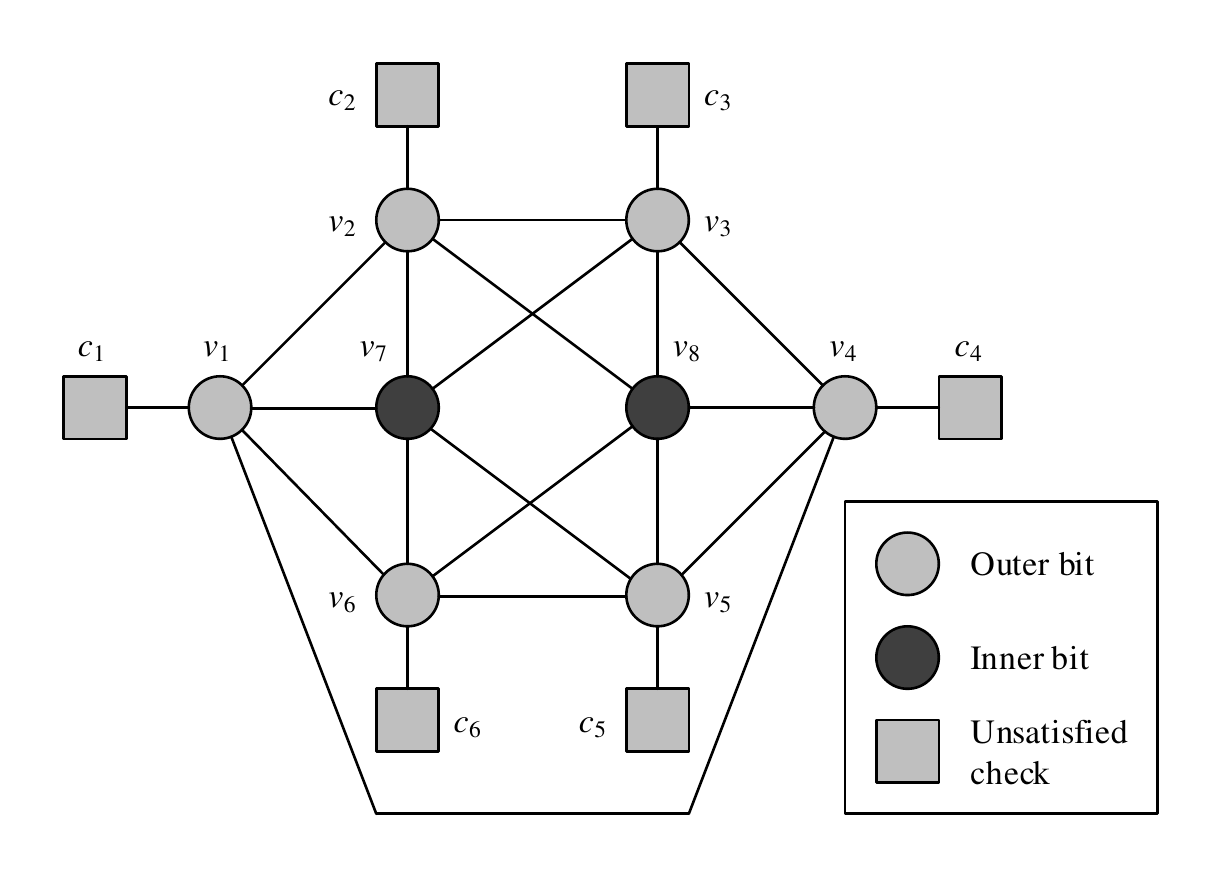}
\caption{Illustration of a type II (8,6) ETS.}
\label{ar_strongset}
\end{figure}

From Table~\ref{tbl:quenchingarray}, one can observe that, unlike the RS-LDPC code discussed above, the error floor of array LDPC code is not dominated by only one kind of ETS error, but attributed to several kinds of ETS errors, including (8,6), (9,5), (8,8), and (10,4) ETS errors \cite{zhang2009design}. An (8,6) ETS is illustrated in \figurename~\ref{ar_strongset}. It is an example of type II ETS. \textbf{A Type II ETS is one in which each VN is connected to no more than 1 degree-1 CN, and at least 1 VN is not connected to any degree-1 CN. The VNs that have no neighboring degree-1 CN are called inner bits, and the VNs that have only 1 neighboring degree-1 CN are called outer bits.}

In the type II (8,6) ETS illustrated in \figurename~\ref{ar_strongset}, the inner bits, $v_7$ and $v_8$, are connected to all satisfied checks, through which they reinforce the outer bits in the ETS. The inner bits are more ``deeply'' trapped than the outer bits since they are not connected to any unsatisfied checks. One iteration of heating helps correct the outer bits, but it does not propagate to the inner bits. The immediate cooling after only one iteration of heating hampers the full recovery. In annealing language, the temperature of the outer bits rise after the heating step, but the inner bits are still cold. Therefore, we propose a second post-processing method called extended heating by setting $P>1$ in Algorithm~\ref{alg_pp}.

Compared to quenching, extended heating prolongs heating to $P$ iterations, where $P>1$, before cooling. The idea is to heat all the bits in the ETS, including both outer and inner bits, to raise the temperature evenly. The neighborhood set is updated after each iteration of heating, allowing the set to be enlarged to include inner bits so that heating can be propagated to them. Prolonged heating allows the bits in a ETS to accumulate enough energy to avoid falling back to the same local minimum.

Among the 236 ETS errors from FPGA emulations, there are only 1 type I ETS error and 184 type II ETS errors that are listed in Table~\ref{tbl:typeiisets}. Quenching with $P$ = 1 and $A_0$ = 1 resolves the type I ETS error but only 84\% of the type II ETS errors. In comparison, extended heating with $P$ = 10 and $A_0$ = 1 resolves 97\% of the type II ETS errors, which demonstrates its effectiveness. When the number of inner bits is large, e.g., in (10,4), (10,6), and (11,5) ETS errors, the success rate of quenching is particularly low, but extended heating works well consistently.

\begin{table}
	\centering
	\caption{Type II ETS Error Profile of the (2209, 1978) Array LDPC Code and Effectiveness of Quenching and Extended Heating}
	\renewcommand{\arraystretch}{1.2}
	\begin{tabular}{|c|c|c|c|c|c|c|}
		\hline
		\multirow{2}{*}{ETS} & Inner & Error & Resolved by & Resolved by\\
		& bits & count & quenching & extended heating\\ \hline\hline
		(8,6)  & 2 & 124  & 105 (85\%) & 121 (98\%)\\ \hline
		(9,5)  & 4 &  37   & 33 (89\%)  &  36 (97\%)\\ \hline
		(10,4) & 6 &  12   &  8 (67\%)   &  12 (100\%)\\ \hline
		(10,6) & 4 &   5   &  3 (60\%)   &   4 (80\%)\\ \hline
		Other & -  &   6   &  6 (100\%)  &  6 (100\%)\\ \hline\hline
		Total  & -  &  184 & 155 (84\%) & 179 (97\%)\\ \hline
	\end{tabular}
	\label{tbl:typeiisets}
\end{table}

\subsection{Type III ETS and Focused Heating}
\label{sec:type3}

\begin{figure}
\centering
\includegraphics[width=.75\linewidth]{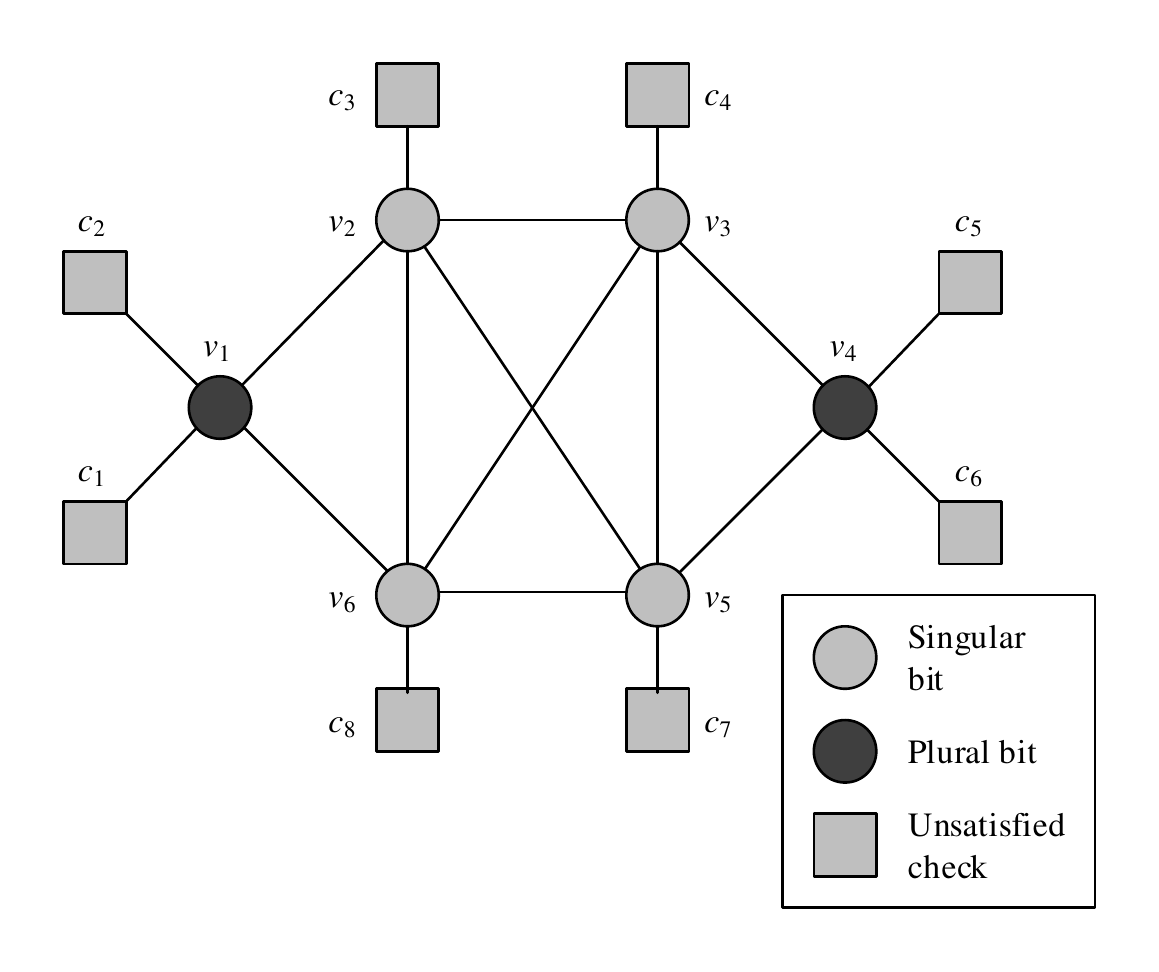}
\caption{Illustration of a type III (6,8) ETS.}
\label{ar_weakset}
\end{figure}

Besides type I and type II ETS errors, there are 58 additional ETS errors collected for the $Q4.0$ (2209, 1978) array LDPC decoder. The (6,8) ETS shown in \figurename~\ref{ar_weakset} is an example of them. The (6,8) ETS is neither type I, nor type II, because two of the bits, $v_1$ and $v_4$ are each connected to two unsatisfied checks. It is an example of type III ETS. \textbf{A Type III ETS is one in which 1 or more VNs are each connected to more than 1 degree-1 CNs. The VNs that have only 1 neighboring degree-1 CN are called singular bits, and the VNs that have more than 1 neighboring degree-1 CNs are called plural bits.}

A type III ETS typically has more unsatisfied checks than the size of the ETS. Since the neighborhood set is traced from the unsatisfied checks, the neighborhood set is relatively larger. A large neighborhood set means more bits, mostly correct bits, are perturbed in heating. Table~\ref{tbl:typeiiisets} above lists the dominant type III ETS errors that have been collected in the error floor region. Quenching with $P$ = 1 and $A_0$ = 1 resolves only 55\% of the type III ETS errors. Extended heating with $P$ = 10 and $A_0$ = 1 resolves 86\%. Overheating is the problem in both cases that cause the two methods to be not as effective.

\begin{table}
	\centering
	\caption{Type III ETS Error Profile of the (2209, 1978) Array LDPC Code and Effectiveness of Extended Heating and Focused Heating}
	\renewcommand{\arraystretch}{1.2}
	\begin{tabular}{|c|c|c|c|c|c|}
		\hline
		\multirow{3}{*}{ETS} & \multirow{3}{*}{\shortstack{Plural\\bits}} & \multirow{3}{*}{\shortstack{Error\\count}} & \multicolumn{3}{c|}{Resolved by}\\ \cline{4-6}
		& 		 & 			 & \multirow{2}{*}{quenching} & extended & extended and \\
		& 		 &       &           & heating  & focused heating \\ \hline\hline
		(6,8)  & 2 &  5 & 3 (60\%) &  4 (80\%) & 5 (100\%) \\ \hline
		(8,8)  & 2 & 11 & 7 (64\%) &  10 (91\%) & 11 (100\%)\\ \hline
		(8,8)  & 1 &  4 & 4 (100\%) &  4 (100\%) & 4 (100\%)\\ \hline
		(8,8)  & 4 &  4 & 1 (25\%) &  4 (100\%) &  4 (100\%)\\ \hline
		(10,8) & 1 &  4 & 4 (100\%) &  4 (100\%) &  4 (100\%)\\ \hline
		other  & - & 30 & 13 (43\%) & 24 (80\%) & 28 (93\%) \\ \hline\hline
		Total  & - & 58 & 32 (55\%) & 50 (86\%) & 56 (97\%) \\ \hline
	\end{tabular}
	\label{tbl:typeiiisets}
\end{table}

In a type III ETS, a plural bit, e.g., $v_1$ or $v_4$ in \figurename~\ref{ar_weakset}, is connected to more than one unsatisfied checks. Therefore, a plural bit candidate can be identified as one that is connected to more than one unsatisfied checks. After the plural bit candidates are identified, they can be corrected by bit flipping, allowing the unsatisfied checks, e.g., $c_1$, $c_2$, $c_5$ and $c_6$ in \figurename~\ref{ar_weakset}, to be turned to satisfied checks. After bit flipping, heating can be applied to a smaller and focused neighborhood set to be more effective.

In this case, bit flipping acts as another form of perturbation. To control the noise injection, we use soft bit flipping, i.e., reduce the reliability of the soft decision to a low value $B_0$ to weaken the plural bits without a significant impact on the correct bits outside the ETS. We call this method focused heating as described in Algorithm~\ref{alg_focheating}.

\begin{postproc}
\label{alg_focheating}
Focused Heating
	\begin{enumerate}
	\item BP decoding: run for $M$ iterations. If there are unsatisfied checks, continue post-processing.
	\item Post-processing:
		\begin{enumerate}
			\item Constraining: run $L$ iterations of soft bit flipping.
			\item Cooling: run $N$ iterations of BP.
		\end{enumerate}
	\end{enumerate}
\end{postproc}

Focused heating uses $L$ iterations of soft bit flipping to selectively weaken the plural bits and shrink the neighborhood set, so that extended heating can be applied to a focused neighborhood set. In practice, extended heating and focused heating need to be combined because a type III ETS error can include inner bits that need to be resolved by extended heating. Assume a gap of $G$ iterations that separates extended and focused heating. Extended and focused heating  with $P$ = 10 and $A_0$ = 1 (for extended heating), $L$ = 5 and $B_0$ = 3 (for focused heating), and $G$ = 10 resolves 97\% of the type III ETS errors, as shown in Table~\ref{tbl:typeiiisets}, more effective than quenching or extended heating.

\begin{figure}
\centering
\includegraphics[width=.75\linewidth]{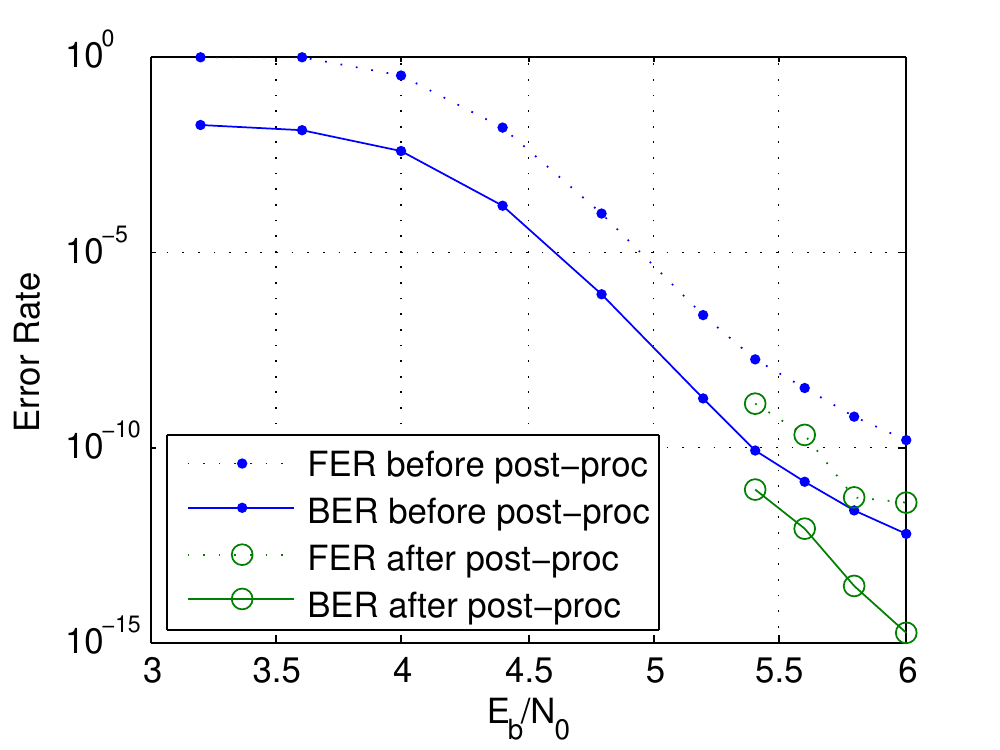}
\caption{Error rate of the (2209, 1978) array LDPC code before and after post-processing using extended and focused heating.}
\label{ar_ber}
\end{figure}

In total, extended and focused heating can be applied to resolve 99\% of the 236 ETS errors collected in the error floor region of the $Q4.0$ (2209, 1978) array LDPC decoder. Note that although the method is designed for ETS errors, extended and focused heating can resolve 82\% of the 38 non-ETS errors. Table~\ref{tbl:ar_results} lists the summary of the results. The BER in the error floor region is lowered by more than 2 orders of magnitude at $E_b/N_0$ = 6.0dB, as shown in \figurename~\ref{ar_ber}. This experiment is done with $P$ = 10 and $A_0$ = 1 (for extended heating), $L$ = 5 and $B_0$ = 3 (for focused heating), and $G$ = 10.

\begin{table}
	\centering
	\caption{Summary of ETS and Non-ETS Errors of the (2209,1978) Array LDPC Decoder in the Error Floor Region and the Effectiveness of Post-Processing by Combined Extended and Focused Heating}
	\renewcommand{\arraystretch}{1.2}
	\begin{tabular}{|c|c|c|c|}
		\hline
		\multirow{2}{*}{$E_b/N_0$} & Error & Number & Resolved by extended\\
		& type & of errors & and focused heating\\ \hline\hline
		\multirow{4}{*}{5.6 dB} & Type I ETS   & 1  & 1 (100\%)\\ \cline{2-4}
		& Type II ETS  & 74 & 73 (99\%)\\ \cline{2-4}
		& Type III ETS & 20 & 19 (95\%)\\ \cline{2-4}
		& Non-ETS      & 27 & 21 (78\%)\\ \hline\hline
		\multirow{4}{*}{5.8 dB} & Type I ETS   & 0  & - \\ \cline{2-4}
		& Type II ETS  & 82 & 82 (100\%)\\ \cline{2-4}
		& Type III ETS & 21 & 20 (95\%)\\ \cline{2-4}
		& Non-ETS      & 9  & 9 (100\%)\\ \hline\hline
		\multirow{4}{*}{6.0 dB} & Type I ETS   & 0  & - \\ \cline{2-4}
		& Type II ETS  & 33 & 33 (100\%)\\ \cline{2-4}
		& Type III ETS & 5  & 5 (100\%)\\ \cline{2-4}
		& Non-ETS      & 2  & 1 (50\%)\\ \hline\hline
		\multirow{2}{*}{Total} & ETS errors   & 236 & 233 (99\%)\\ \cline{2-4}
		& Non-ETS errors & 38 & 31 (82\%)\\ \hline
	\end{tabular}
	\label{tbl:ar_results}
\end{table}

\section{Application of Post-Processing Methods -- Case Study on an IEEE 802.11n LDPC Code}

\begin{figure*}
\centering
\includegraphics[width=.75\linewidth]{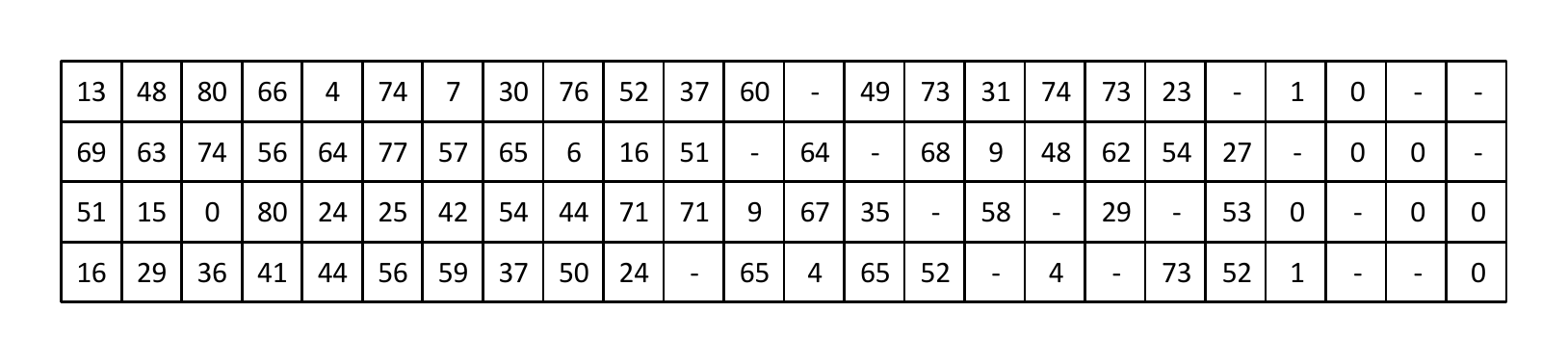}
\caption{Parity check matrix of the (1944, 1620) LDPC code for IEEE 802.11n standard \cite{std11n2009}.}
\label{11n_matrix}
\end{figure*}

The focused and extended heating methods are developed based on the (2209, 1978) array LDPC code, but the methods are generally applicable. We demonstrate these methods on an arbitrarily selected rate-0.83 (1944, 1620) LDPC code for the IEEE 802.11n standard \cite{std11n2009}. The $H$ matrix of the (1944, 1620) LDPC code is made up of a 4$\times$24 array of 81$\times$81 identity matrices, cyclic shifted identity matrices, or zero matrices. The $H$ matrix is described in \figurename~\ref{11n_matrix} \cite{std11n2009}, where a ``0'' indicates an 81$\times$81 identity matrix, a number $x$, $x>0$, indicates an 81$\times$81 matrix obtained by right cyclic shifting of the identity matrix by $x$, and a ``-'' indicates an 81$\times$81 zero matrix.

The (1944, 1620) LDPC code is structured but not regular. Note that the identity matrices are laid out in a staircase on the right hand side to allow for an efficient encoder design. This design however leads to a low minimum column degree of 2. which dictates the majority of the error patterns.

\begin{figure}
\centering
\includegraphics[width=.75\linewidth]{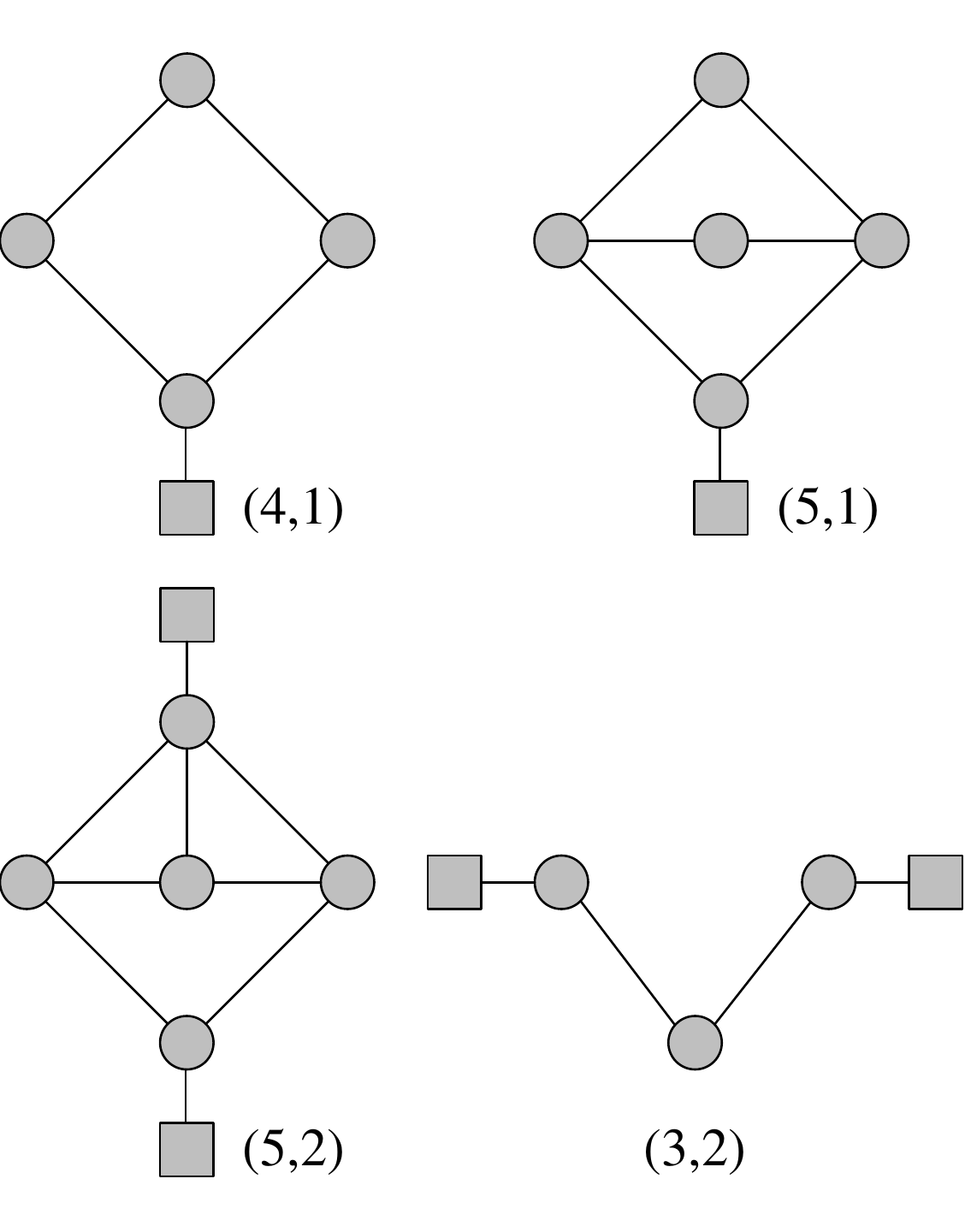}
\caption{Dominant type II ETS structures in the (1944, 1620) LDPC code for the IEEE 802.11n standard.}
\label{11n_sets}
\end{figure}

\begin{figure}
\centering
\includegraphics[width=.75\linewidth]{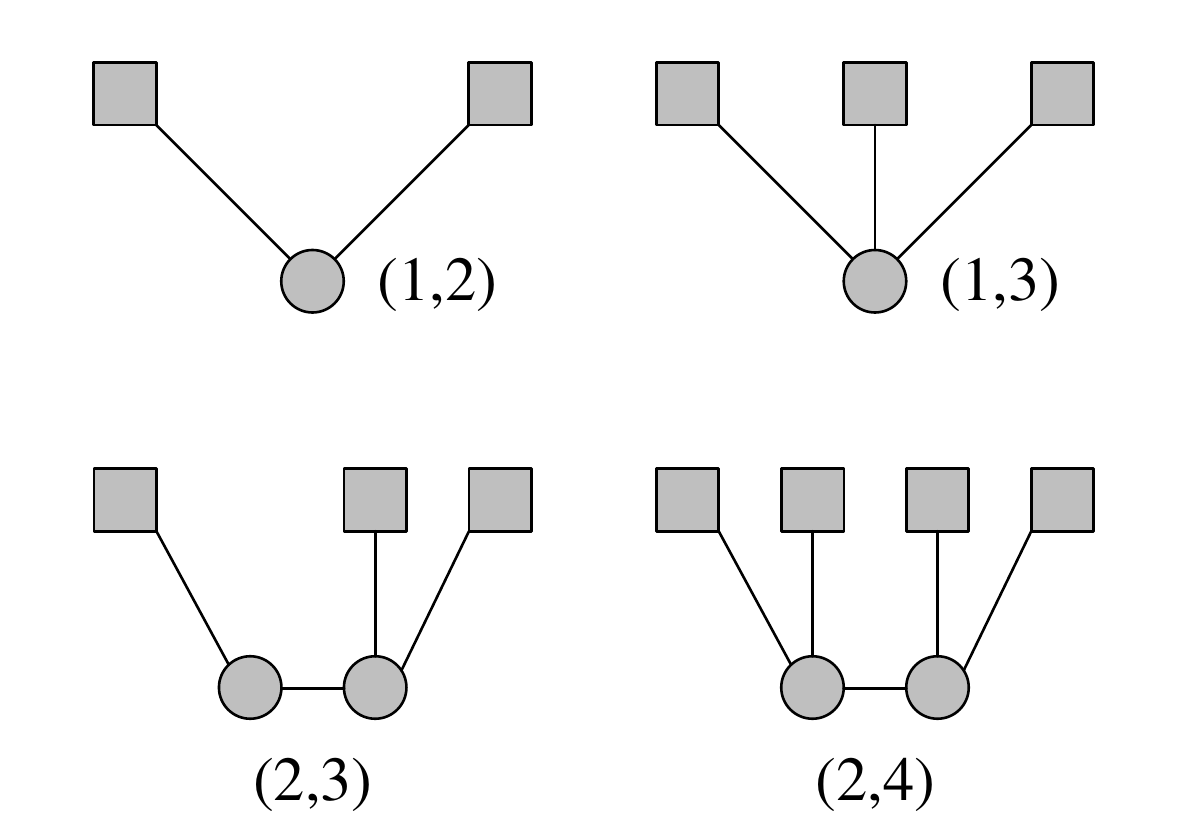}
\caption{Dominant type III ETS structures in the (1944, 1620) LDPC code for IEEE 802.11n standard.}
\label{11n_non_as}
\end{figure}

We implemented a $Q5.0$ decoder for this LDPC code on FPGA and collected 1830 errors in the error floor region ($E_b/N_0$ = 5.0 dB, 5.4 dB, 5.6 dB, and 5.8 dB). More than 99\% of the errors are ETS errors, and the remaining are non-ETS errors. Type I, type II and type III account for 7.3\%, 70\% and 22.5\% of the ETS errors, respectively. The dominant type II ETS errors, (3,2), (4,1), (5,1), and (5,2), account for 76\% of the type II ETS errors, and their structures are illustrated in \figurename~\ref{11n_sets}. Since they all contain inner bits, extended heating can be applied. The dominant type III ETS errors, (1,2), (1,3), (2,3) and (2,4), account for 82\% of the type III ETS errors. Their structures are illustrated in \figurename~\ref{11n_non_as}. Extended and focused heating are applicable to these errors.

\begin{figure}
\centering
\includegraphics[width=.75\linewidth]{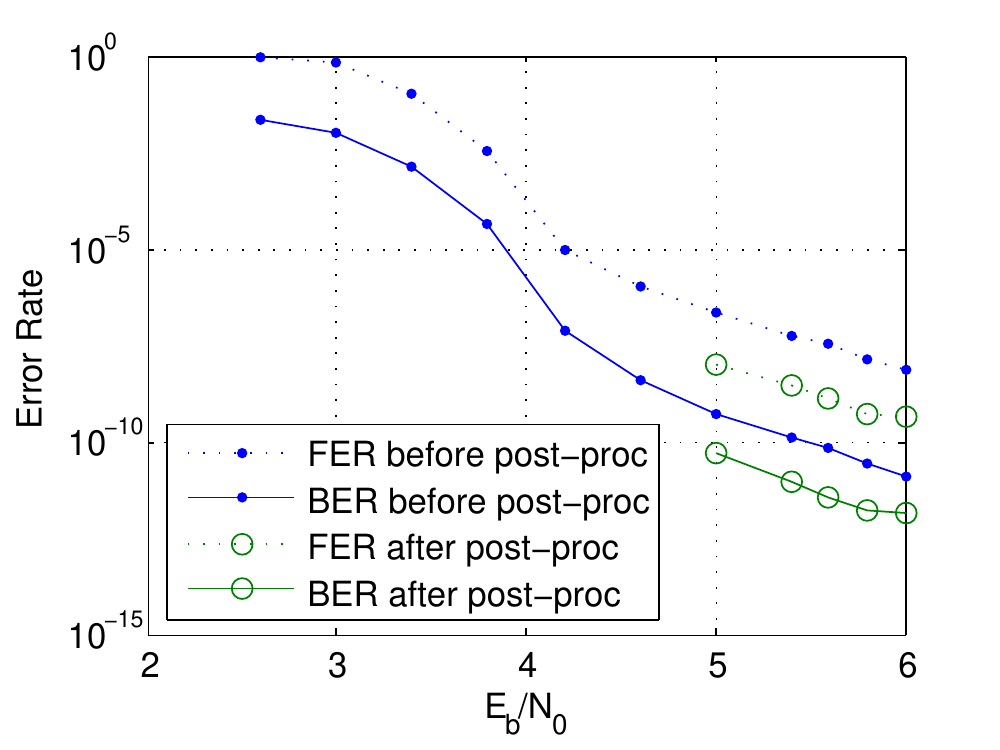}
\caption{Error rate of the (1944, 1620) IEEE 802.11n LDPC code before and after post-processing using extended and focused heating.}
\label{11n_ber}
\end{figure}

Extended and focused heating is used to post-process the errors collected in the $Q5.0$ (1944, 1620) IEEE 802.11n LDPC decoder. Table~\ref{tbl:ar_results_802} shows that the overall success rate is 95\%. The error floor is reduced by one to two orders of magnitude after post-processing, as shown in \figurename~\ref{11n_ber}. The results are obtained with $P$ = 10 and $A_0$ = 1 (for extended heating), $L$ = 5 and $B_0$ = 1 (for focused heating), and $G$ = 10. In comparison, quenching alone \cite{zhang2008lowering} resolves 23\% of the errors and the bi-mode syndrome erasure decoding \cite{han2009low} resolves 59\% of the errors based on our simulations.

\begin{table}
	\centering
	\caption{Summary of ETS and Non-ETS Errors of the (1944,1620) IEEE 802.11n LDPC Decoder in the Error Floor Region and the Effectiveness of Post-Processing by Combined Extended and Focused Heating}
	\renewcommand{\arraystretch}{1.2}
	\begin{tabular}{|c|c|c|c|}
		\hline
		\multirow{2}{*}{$E_b/N_0$} & Error & Number & Resolved by extended\\
		& type & of errors & and focused heating\\ \hline\hline
		\multirow{4}{*}{5.0 dB} & type I ETS & 34 & 31 (91\%)\\ \cline{2-4}
		& type II ETS  & 351 & 338 (96\%)\\ \cline{2-4}
		& type III ETS & 91 & 86 (95\%)\\ \cline{2-4}
		& non-ETS      & 4 & 4 (100\%)\\ \hline\hline
		\multirow{4}{*}{5.4 dB} & type I ETS & 26 & 23 (88\%)\\ \cline{2-4}
		& type II ETS  & 324 & 311 (96\%)\\ \cline{2-4}
		& type III ETS & 100 & 94 (94\%)\\ \cline{2-4}
		& non-ETS      & 1 & 1 (100\%)\\ \hline\hline
		\multirow{4}{*}{5.6 dB} & type I ETS & 38 & 37 (97\%)\\ \cline{2-4}
		& type II ETS  & 319 & 306 (96\%)\\ \cline{2-4}
		& type III ETS & 102 & 96 (94\%)\\ \cline{2-4}
		& non-ETS      & 2 & 1 (50\%)\\ \hline\hline
		\multirow{4}{*}{5.8 dB} & type I ETS & 35 & 32 (91\%)\\ \cline{2-4}
		& type II ETS  & 284 & 267 (94\%)\\ \cline{2-4}
		& type III ETS & 119 & 113 (95\%)\\ \cline{2-4}
		& non-ETS      & 0 & 0 (N/A)\\ \hline\hline
		\multirow{4}{*}{Total} & type I ETS & 133 & 123 (92\%)\\ \cline{2-4}
		& type II ETS  & 1278 & 1222 (96\%)\\ \cline{2-4}
		& type III ETS & 412 & 389 (94\%)\\ \cline{2-4}
		& non-ETS      & 7 & 6 (86\%)\\ \hline
	\end{tabular}
	\label{tbl:ar_results_802}
\end{table}

The device utilization of a row-parallel 802.11n LDPC decoder is listed in Table~\ref{tbl:hardware}. The addition of extended and focused heating introduces less than 10\% overhead. The results of this work are compared in Table~\ref{tbl:hardware_comparison} with two prior designs \cite{chen2011hardware,zhang2010efficient} that included hardware design and evaluation. This work demonstrates a lower datapath overhead than \cite{zhang2010efficient} and requires significantly less memory than \cite{chen2011hardware}. As a deterministic method, this work features a lower latency than \cite{chen2011hardware} because it is integrated as part of BP decoding, while \cite{chen2011hardware} requires trial and error.

\begin{table}
	\centering
	\caption{Device Utilization of 4-Row-Parallel IEEE 802.11n (1944, 1620) decoders (based on Xilinx Virtex-5 XC5VLX155T FPGA)}
	\renewcommand{\arraystretch}{1.2}
	\begin{tabular}{|c|c|c|c|c|c|}
		\hline
		\multirow{2}{*}{Design} & \multirow{2}{*}{Baseline} & \multirow{2}{*}{Quenching} & Extended + focused \\
		& & & heating\\ \hline\hline
		Slice & 4,432 & 4,611 & 4,633 \\ 
		registers & (4.60\%) & (4.79\%) & (4.81\%) \\\hline
		Slice & 10,066 & 10,732 & 10,901\\ 
		LUTs & (10.4\%) & (11.1\%) & (11.2\%) \\\hline
		Occupied & 4,782 & 4,834 & 4,844 \\ 
		slices & (19.7\%) & (19.9\%) & (19.9\%) \\\hline
		\multirow{2}{*}{BRAMs}  & 35 & 35 & 35 \\
		& (16.4\%) & (16.4\%) & (16.4\%) \\\hline
	\end{tabular}
	\label{tbl:hardware}
\end{table}

\begin{table}
	\centering
	\caption{Comparison of Low-Floor LDPC Decoder Implementations}
	\renewcommand{\arraystretch}{1.2}
	\begin{tabular}{|c|c|c|c|}
		\hline
		& This work & [20] & [28] \\ \hline\hline
		Implementation & FPGA & Synthesis & Silicon \\ \hline
		\multirow{2}{*}{Method} & Generalized & \multirow{2}{*}{Backtracking} & \multirow{2}{*}{Quenching} \\ 
		& post-processing & & \\\hline
		\multirow{2}{*}{Code} & \multirow{2}{*}{Any} & \multirow{2}{*}{Any} & (2048,1723)\\ 
		& & & RS-LDPC \\\hline
		Datapath & \multirow{2}{*}{8.3\%} & \multirow{2}{*}{7\%} & \multirow{2}{*}{13.7\%} \\
		overhead & & & \\ \hline
		Memory & \multirow{2}{*}{4.5\%} & \multirow{2}{*}{46\%} & \multirow{2}{*}{N/A} \\ 
		overhead & & & \\\hline
	\end{tabular}
	\label{tbl:hardware_comparison}
\end{table}

\section{Conclusions}

Error floors of structured LDPC codes are caused by local minima due to non-codeword ETS and ETS-like errors. Inspired by simulated annealing, we design post-processing methods to perturb the local minimum state, followed by cooling to help decoding converge to the global minimum.

We use three well-known LDPC code examples, a (2048, 1723) RS-LDPC code, a (2209, 1978) array LDPC code, and a (1944, 1620) 802.11n LDPC code for the IEEE 802.11n standard to demonstrate three types of ETS structures: type I with one-to-one correspondence between each unsatsified check and ETS bit, type II with inner bits, i.e., ETS bits that are not connected to any unsatisfied check, and type III with plural bits, i.e., ETS bits that are connected to more than one unsatisfied checks.

Three post-processing methods are proposed to resolve ETS errors. The quenching algorithm uses one heating step followed by immediate cooling to resolve type I ETS errors. The extended heating algorithm prolongs heating to multiple steps to allow the inner bits to accumulate enough energy to resolve type II ETS errors. The focused heating algorithm applies soft bit flipping to the plural bits in order to correct them and narrow down the neighborhood set for more effective heating. The post-processing parameters used in this work are summarized in Table~\ref{tbl:pp_parameters}.

\begin{table}
	\centering
	\caption{Summary of Post-Processing Parameters}
	\renewcommand{\arraystretch}{1.2}
	\begin{tabular}{|c|c|c|c|c|c|c|}
		\hline
		Code & Method & $P$ & $A_0$ & $L$ & $B_0$ & $G$ \\ \hline\hline
		(2048,1723) & \multirow{2}{*}{Quenching} & \multirow{2}{*}{1} & \multirow{2}{*}{1} & \multirow{2}{*}{-} & \multirow{2}{*}{-} &\multirow{2}{*}{-} \\ 
		RS-LDPC & & & & & & \\\hline
		(2209, 1978) & Extended + & \multirow{2}{*}{10} & \multirow{2}{*}{1} & \multirow{2}{*}{5} & \multirow{2}{*}{3} & \multirow{2}{*}{10}\\ 
		array LDPC & focused heating & & & & & \\ \hline
	    (1944, 1620) & Extended + & \multirow{2}{*}{10} & \multirow{2}{*}{1} & \multirow{2}{*}{5} & \multirow{2}{*}{1} & \multirow{2}{*}{10} \\ 
		802.11n LDPC & focused heating & & & & & \\ \hline
	\end{tabular}
	\label{tbl:pp_parameters}
\end{table}

The post-processing methods can be easily integrated as part of BP decoding, adding minimal overhead to the hardware implementation. As these methods are conditionally triggered when the decoder fails to converge at a very low BER level, the impact on decoding throughput and energy consumption is negligible.

The methods are demonstrated by post-processing the errors collected in the error floor region of the three LDPC code examples. The success rate is over 95\% for ETS errors and over 80\% for non-ETS errors for the IEEE 802.11n (1944,1620) LDPC code.

\bibliographystyle{IEEEtran}
\bibliography{IEEEabrv,tcsi18}

\begin{IEEEbiography}[{\includegraphics[width=1in,height=1.25in,clip,keepaspectratio]{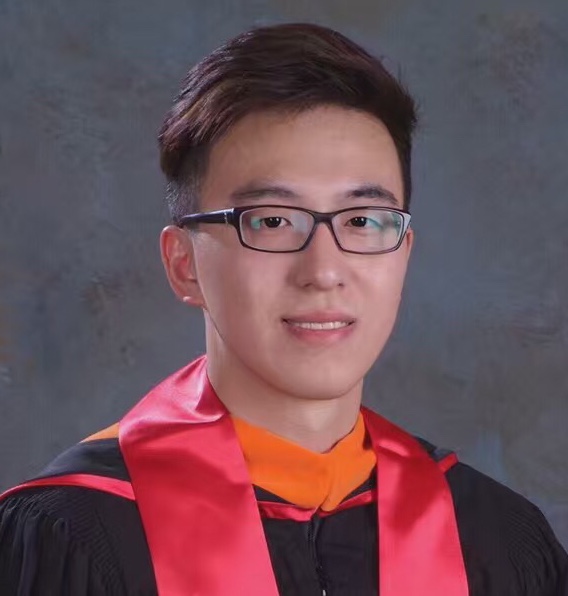}}]{Yaoyu Tao}
	(S'11) received the B.S. degree in electrical and computer engineering from Shanghai Jiao Tong University, Shanghai, China, in 2011, the B.S. degree in electrical engineering from the University of Michigan, Ann Arbor, MI, USA, in 2013, and the M.S. degree in electrical engineering from Stanford University, Stanford, CA, USA, in 2015. He is currently with Qualcomm Wireless R\&D, San Jose, CA as a senior research engineer and also pursuing his Ph.D. degree. His research interests are in high-speed energy efficient architecture design for wireless link, especially for MIMO detection and channel coding, and high-performance VLSI systems design for machine-learning applications.
\end{IEEEbiography}

\begin{IEEEbiography}[{\includegraphics[width=1in,height=1.25in,clip,keepaspectratio]{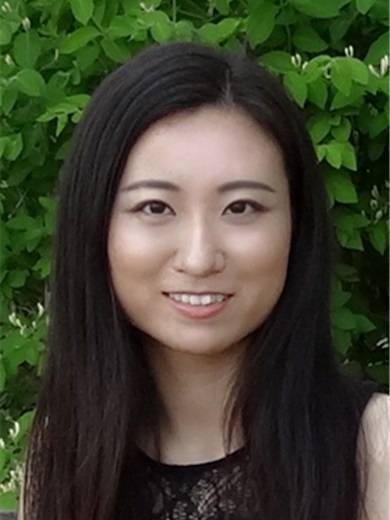}}]{Shuanghong Sun}
    (S'11--M'17) received the B.S. degree in electrical and computer engineering from Shanghai Jiao Tong University, Shanghai, China in 2012, and the B.S., M.S. and Ph.D. degrees in electrical engineering from the University of Michigan, Ann Arbor, MI, USA, in 2012, 2014, and 2017, respectively. She is currently with Intel Corp., San Jose, CA as a DSP algorithm engineer.

    Dr. Sun was with Broadcom Corp., Irvine, CA and Qualcomm Inc., San Diego, CA in 2015. Her research interests are channel coding, digital architectures and high-performance VLSI systems.
\end{IEEEbiography}

\begin{IEEEbiography}[{\includegraphics[width=1in,height=1.25in,clip,keepaspectratio]{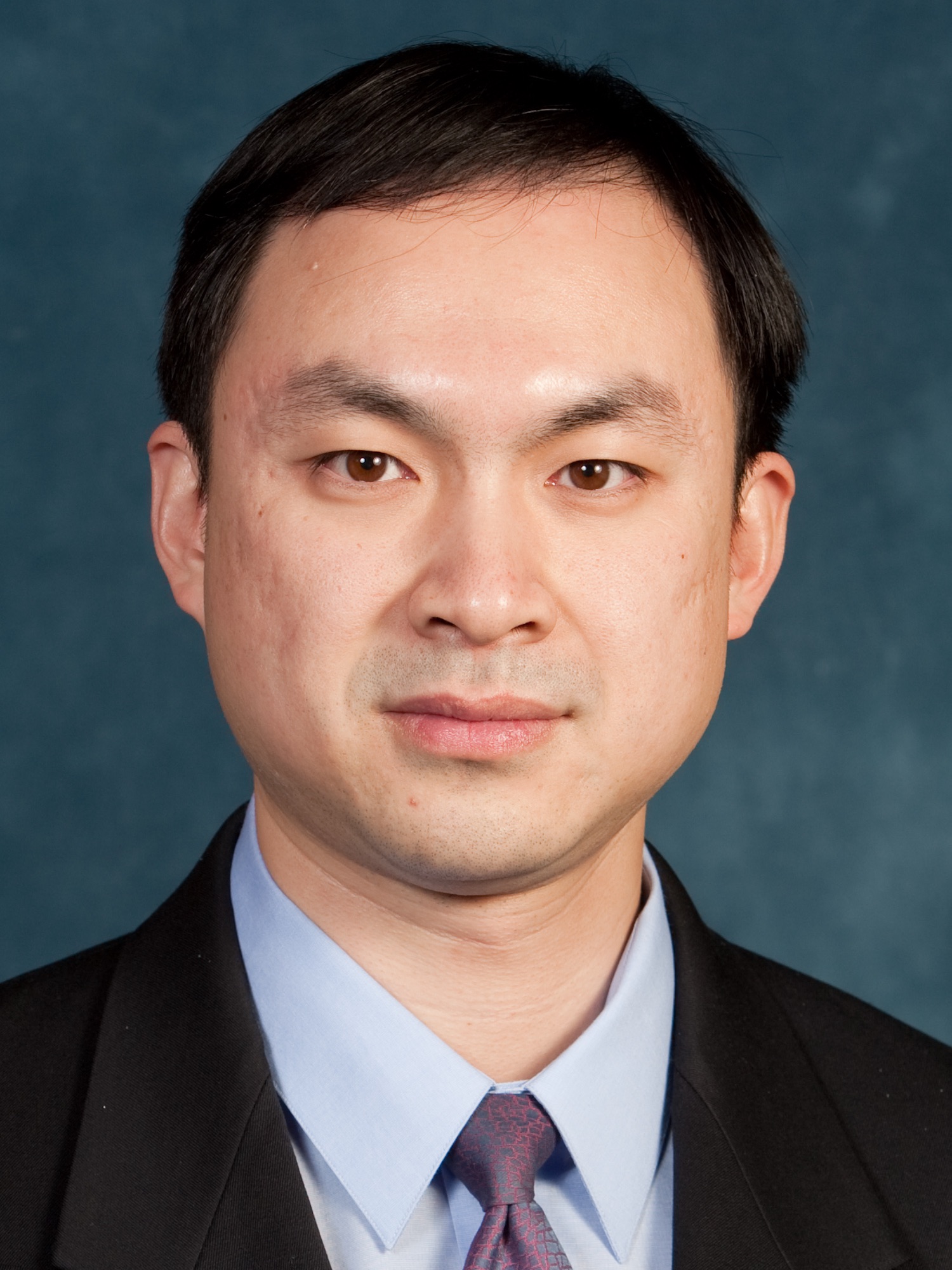}}]{Zhengya Zhang}
    (S'02--M'09--SM'17) received the B.A.Sc. degree in computer engineering from the University of Waterloo, Ontario, Canada, in 2003, and the M.S. and Ph.D. degrees in electrical engineering from the University of California, Berkeley (UC Berkeley), in 2005 and 2009, respectively. He has been a faculty member with the University of Michigan, Ann Arbor, since 2009, where he is currently an Associate Professor with the Department of Electrical Engineering and Computer Science. His current research interests include low-power and high-performance VLSI circuits and systems for computing, communications, and signal processing.
    
    Dr. Zhang was a recipient of the David J. Sakrison Memorial Prize from UC Berkeley in 2009, the National Science Foundation CAREER Award in 2011, the Intel Early Career Faculty Award in 2013. He has been an Associate Editor of the IEEE TRANSACTIONS ON VERY LARGE SCALE INTEGRATION SYSTEMS since 2015. He serves on the Technical Program Committees of Symposium on VLSI Circuits and IEEE Custom Integrated Circuits Conference (CICC). He was an Associate Editor of the IEEE TRANSACTIONS ON CIRCUITS AND SYSTEMS—PART I: REGULAR PAPERS (2013-2015) and the IEEE TRANSACTIONS ON CIRCUITS AND SYSTEMS—PART II: EXPRESS BRIEFS (2014-2015).
\end{IEEEbiography}

\end{document}